\documentclass[prd,
,superscriptaddress,showpacs,nofootinbib,%
tightenlines ]{revtex4}
\usepackage{epsfig}
\usepackage{amssymb,color}
\usepackage{amsfonts}
\usepackage{amssymb}
\usepackage{slashed}

\newcommand{\ben}{\begin{displaymath}}
\newcommand{\een}{\end{displaymath}}
\newcommand{\be}{\begin{equation}}
\newcommand{\ee}{\end{equation}}
\newcommand{\bea}{\begin{eqnarray}}
\newcommand{\eea}{\end{eqnarray}}
\begin{document}
\title{Vacuum energy in the effective field theory of general relativity II:\\
  Inclusion of fermions and a comment on the QCD contribution}

\author{J.~Gegelia}
\affiliation{Ruhr-University Bochum, Faculty of Physics and Astronomy,
Institute for Theoretical Physics II, D-44870 Bochum, Germany}
\affiliation{Tbilisi State  University,  0186 Tbilisi,
 Georgia}
\author{Ulf-G.~Mei{\ss}ner}
 \affiliation{Helmholtz Institut f\"ur Strahlen- und Kernphysik and Bethe
   Center for Theoretical Physics, Universit\"at Bonn, D-53115 Bonn, Germany}
 \affiliation{Institute for Advanced Simulation, Institut f\"ur Kernphysik
   and J\"ulich Center for Hadron Physics, Forschungszentrum J\"ulich, D-52425 J\"ulich,
Germany}
\affiliation{Tbilisi State  University,  0186 Tbilisi, Georgia}

\begin{abstract}
Recently in the framework of a two-loop order calculation for an effective field theory of
scalar and vector fields interacting with the metric field we have shown 
that for the cosmological constant term which is fixed by the condition of vanishing
vacuum energy the graviton remains massless and
there exists a self-consistent effective field theory of general relativity defined on a
flat Minkowski background. In the current paper we extend the two-loop  analysis for
an effective field theory of fermions interacting with the
gravitational field and obtain an analogous result.
We also address the issues of fine tuning of the strong interaction contribution to the vacuum
energy and the  compatibility  of chiral symmetry in the light quark sector with the consistency of
the effective field theory of general relativity in  a flat Minkowski background. 

\end{abstract}


\maketitle

\section{Introduction}	
	
It is widely accepted that at low energies the  physics of the fundamental particles can be
adequately described by effective field theories (EFTs), with the Standard Model being its leading
order approximation \cite{Weinberg:1995mt}. 
Gravitation can also be included in this framework by considering the effective Lagrangian of metric
fields interacting with matter fields  \cite{Donoghue:1994dn,Donoghue:2015hwa}. 
Within this approach the metric field is represented as the Minkowski
background plus the graviton field and the cosmological constant is usually set equal to zero,
see, e.g., Ref.~\cite{Donoghue:2017pgk}.
For a non-vanishing cosmological constant term $\Lambda$ the graviton propagator has a pole corresponding
to  a massive ghost mode \cite{Veltman:1975vx}. 
As the cosmological constant term is not suppressed by any symmetry of the effective 
theory, setting it to zero does not solve
the problem, because the radiative corrections re-generate the massive ghost \cite{Burns:2014bva}.   
It has been shown in Ref.~\cite{Burns:2014bva} that one can represent the cosmological constant as a
power series in $\hbar$ and adjust the  coefficients of this series such that the unphysical mass
of the graviton is cancelled to all orders in the loop expansion.  
Thus,  to take into account  a cosmological constant term other than obtained in 
Ref.~\cite{Burns:2014bva} it is necessary to consider an EFT in a curved background field.
As shown in Ref.~\cite{Gabadadze:2003jq} by
imposing the equations of motion with respect to the non-trivial background 
graviton field, the mass term of the graviton is removed  at tree level. 
A systematic study of the issue by including the quantum corrections requires an 
EFT on a curved background metric which, to the best of our knowledge, is not available yet. 

The accelerating expansion of the universe (see, e.g., Ref.~\cite{Rubin:2016iqe} and
references therein) leaves us with a huge discrepancy between the measured small value of
the effective cosmological constant and its theoretical estimation \cite{Weinberg:1988cp}. 
In our opinion if there exists any condition that uniquely fixes the value of the cosmological constant,
then it is most natural to expect  that it must be imposed by demanding that the energy of the
physical vacuum state of the theory describing the universe is exactly zero. 
In our recent work~\cite{Gegelia:2019fjx} we calculated two-loop order contributions of a scalar
and vector fields to the vacuum expectation value of the full four-momentum 
in a simplified version of the Abelian model with spontaneous symmetry breaking,
considered also in Ref.~\cite{Burns:2014bva}. 
We found that as a result of a non-trivial cancellation between different diagrams the
requirement of vanishing vacuum energy leads to consistency conditions of the considered EFT, 
first obtained in Ref.~\cite{Burns:2014bva}. In the current work we extend this two-loop analysis
and calculate  the contributions from fermions. 
For the energy-momentum tensor of the gravitational field  we use the definition of the
energy-momentum pseudotensor and the full
four-momentum given in the classic textbook by Landau and Lifshitz~\cite{Landau:1982dva}. 

Further, in the light of the above discussion we re-address the issue of the fine tuning of
the strong interaction contribution to the vacuum energy and 
compatibility of the results of Ref.~\cite{Burns:2014bva} with the 
chiral symmetry of quantum chromodynamics (QCD).  

Our work is orginized as follows:  
In section~\ref{VEC} we specify the details of the considered EFT of fermions interacting
with a gravitational field and calculate one- and two-loop contributions to the vacuum energy
and the vacuum expectation value of the gravitational field. 
In section~\ref{QCDC} we discuss the QCD contribution to vacuum energy and the problem
of the fine-tuning following Ref.~\cite{Donoghue:2016tjk}. 
We summarize in section~\ref{summary} and the appendix contains the Feynman rules involving fermion
fields and  two-loop integrals required in our calculations.

\section{Vacuum energy in an EFT of fermions interacting with gravitons on a Minkowski background}
\label{VEC}

Effective field theory of  matter interacting with  gravity is described by the most general effective
Lagrangian of gravitational and matter fields, which is invariant under 
general coordinate transformations and the other underlying symmetries,
\begin{eqnarray}
S &=& \int d^4x \sqrt{-g}\, \left\{ {\cal L}_{\rm gr}(g)
+{\cal L}_{\rm m}(g,\psi)\right\} \nonumber\\
&=& \int d^4x \sqrt{-g}\, \left\{ \frac{2}{\kappa^2}
(R-2\Lambda)+{\cal L}_{\rm gr,ho}(g)
+{\cal L}_{\rm m}(g,\psi)\right\} \nonumber\\
&=& S_{\rm gr}(g)+S_{\rm m}(g,\psi) ,
\label{action}
\end{eqnarray}
where $\kappa^2=32 \pi G$, with Newton's constant $G=6.70881\cdot 10^{-39}$ ${\rm GeV}^{-2}$,
 $\psi$ and $g^{\mu\nu}$ denote the matter and metric fields, respectively,  $g=\det g^{\mu\nu}$, 
$\Lambda$ is the cosmological constant and $R$ denotes the scalar curvature. Further,
${\cal L}_{\rm m}(g,\psi)$ is the effective Lagrangian of the matter fields interacting with gravity.
Experimental evidence suggests that self-interaction terms of the gravitational field with
higher orders of derivatives,  represented by ${\cal L}_{\rm gr,ho}(g)$, as well as the 
non-renormalizable interactions of ${\cal L}_{\rm m}(g,\psi)$ give contributions to physical
quantities which  are heavily  suppressed for energies accessible by current accelerators.
Vielbein tetrad fields have to be introduced for an EFT with fermions.

To be specific,  consider the action of the fermions interacting with the gravitational field given by
\begin{eqnarray}
S_{\rm m} & = & \int d^4x \sqrt{-g}\, \left\{
\frac{1}{2} \, \bar\psi \, i e^\mu_a\gamma^a \nabla_\mu \psi -\frac{1}{2} \, \nabla_\mu \bar\psi
\, i e^\mu_a\gamma^a\psi  -m \bar\psi\psi  \right\} +{\cal L}_{\rm HO} ,
\label{MAction}
\end{eqnarray}
where  
${\cal L}_{\rm HO}$
denotes the  interactions of higher order, the specific form of which is not important for the
current work as we will not  include them in our calculations. 
The covariant derivative acting on the fermion field has the form 
\begin{eqnarray}
\nabla_\mu \psi &=& \partial_\mu\psi - \omega^{ab}_\mu \sigma_{ab} \psi , \nonumber\\
\nabla_\mu \bar\psi &=& \partial_\mu\bar\psi + \bar\psi \, \sigma_{ab} \, \omega^{ab}_\mu ,
\label{CovD}
\end{eqnarray}
where $\sigma_{ab}=\frac{1}{4}[\gamma_a, \gamma_b]$ and 
\begin{eqnarray}
\omega_\mu^{ab} &=& -g^{\nu\lambda} e^a_\lambda \left( \partial_\mu e_\nu^b
- e^b_\sigma \Gamma^\sigma_{\mu \nu} \right),\nonumber\\
\Gamma^\lambda_{\alpha \beta} &=& \frac{1}{2}\,g^{\lambda\sigma} \left( \partial_\alpha g_{\beta\sigma}
+ \partial_\beta g_{\alpha\sigma} -  \partial_\sigma g_{\alpha\beta} \right)~.
\label{omega}
\end{eqnarray}
The vielbein fields satisfy the following relations:
\begin{eqnarray}
&& e^a_\mu e^b_\nu \eta_{ab}=g_{\mu\nu}, \ \ \  e_a^\mu e_b^\nu \eta^{ab}=g^{\mu\nu}, \nonumber\\
&& e^a_\mu e^b_\nu g^{\mu\nu}=g_{ab}, \ \ \  e_a^\mu e_b^\nu g_{\mu\nu}=g^{ab}.
\label{VRels}
\end{eqnarray}
The energy-momentum tensor corresponding to Eq.~(\ref{MAction}) has the form \cite{Birrell:1982ix}:
\begin{eqnarray}
T_m^{\mu\nu} & = &  
\frac{i}{4} \, \left( \bar\psi \,  e_{a\mu}\gamma^a \nabla_\nu \psi + \bar\psi \,
e_{a \nu}\gamma^a \nabla_\mu \psi  - \nabla_\mu \bar\psi \, e_{a \nu}\gamma^a\psi
-\nabla_\nu \bar\psi \, e_{a\mu}\gamma^a\psi   \right) +T_{\rm HO}^{\mu\nu} ,
\label{MEMT}
\end{eqnarray}
where $T_{\rm HO}^{\mu\nu}$ corresponds to ${\cal L}_{\rm HO}$. Note that we consider one fermion field
with mass $m$, the extension to more fermion fields with equal or different masses is straightforward.

For  the gravitational field we have 
\begin{eqnarray}
T^{\mu\nu}_{\rm gr} (g)&=& 
\frac{4}{\kappa^2} \, \Lambda\,g^{\mu\nu} +  T_{LL}^{\mu\nu}(g)\,,
\label{defTs}
\end{eqnarray}
\noindent
where the pseudotensor $T_{LL}^{\mu\nu}(g)$ is defined via \cite{Landau:1982dva}
\begin{eqnarray}
(-g)T^{\mu\nu}_{LL} (g) &=& \frac{2}{\kappa^2} \left(\frac{1}{8} \, g^{\lambda \sigma } g^{\mu \nu } g_{\alpha \gamma}
g_{\beta\delta} \, \mathfrak{g}^{\alpha \gamma},_{\sigma }
   \, \mathfrak{g}^{\beta \delta},_\lambda -\frac{1}{4} \, g^{\mu \lambda } g^{\nu\sigma } g_{\alpha ,\gamma} g_{\beta \delta }
  \, \mathfrak{g}^{\alpha\gamma},_\sigma \, \mathfrak{g}^{\beta\delta},_\lambda -\frac{1}{4} \, g^{\lambda \sigma } g^{\mu \nu } 
   g_{\beta \alpha} g_{\gamma \delta} \, \mathfrak{g}^{\alpha \gamma},_\sigma \, \mathfrak{g}^{\beta \delta},_\lambda \right.\nonumber\\
&+& \left. \frac{1}{2}\,  g^{\mu \lambda } g^{\nu\sigma } g_{\beta \alpha} g_{\gamma \delta } \, \mathfrak{g}^{\alpha \gamma},_\sigma \, \mathfrak{g}^{\beta\delta},_\lambda 
+g^{\beta \alpha } g_{\lambda \sigma }
  \, \mathfrak{g}^{\nu \sigma},_\alpha \, \mathfrak{g}^{ \mu\lambda},_\beta +\frac{1}{2} \, g^{\mu \nu } g_{\lambda \sigma }
   \, \mathfrak{g}^{\lambda \beta},_\alpha \, \mathfrak{g}^{\alpha\sigma},_\beta \right.\nonumber\\
&-& \left.g^{\mu \lambda } g_{\sigma \beta }
   \, \mathfrak{g}^{\nu \beta},_\alpha \, \mathfrak{g}^{\sigma\alpha},_\lambda 
   -g^{\nu \lambda } g_{\sigma \beta} \, \mathfrak{g}^{\mu\beta},_\alpha \, \mathfrak{g}^{\sigma \alpha},_\lambda
   +\, \mathfrak{g}^{\lambda \sigma},_\sigma \, \mathfrak{g}^{\mu\nu},_\lambda 
   - \, \mathfrak{g}^{\mu \lambda},_\lambda \, \mathfrak{g}^{\nu \sigma},_\sigma \right),
   \label{LLEMT}
\end{eqnarray}
with $\mathfrak{g}^{\mu\nu}=\sqrt{-g} \, g^{\mu\nu}$ and
$\mathfrak{g}^{\mu\nu},_\lambda=\partial\mathfrak{g}^{\mu\nu}/\partial x^\lambda $.

From the full energy-momentum tensor $T^{\mu\nu}=T^{\mu\nu}_{\rm m} (g,\psi)+T^{\mu\nu}_{\rm gr} (g)$
we obtain the conserved full four-momentum of the matter and the gravitational field as~\cite{Landau:1982dva}
\begin{equation}
P^\mu= \int (-g) \, T^{\mu\nu} d S_\nu\,,
\label{EMV}
\end{equation}
where the integration over any hypersurface containing the whole three-dimensional  space is implied. 
Thus,  by demanding that the vacuum expectation value of the energy-momentum tensor times $(-g)$
vanishes, we will obtain a vanishing energy of the vacuum.
The mentioned vacuum expectation value is given by the following path integral:
\begin{eqnarray}
\langle 0| (-g) T^{\mu\nu} |0\rangle &=& 
\int {\cal D }g\, {\cal D}\psi  \, (-g)\left[ T^{\mu\nu}_{\rm gr}(g)
+ T_{\rm m}^{\mu\nu}(g,\psi) \right] \exp\left\{ i \int d^4 x \, \sqrt{-g}\,\left[ {\cal L}_{\rm gr}(g)
+{\cal L}_{\rm m}(g,\psi) +{\cal L}_{\rm GF}\right] \right\},
\label{VacuumE}
\end{eqnarray}
where we have added the gauge fixing term 
\begin{equation}
{\cal L}_{\rm GF}= \xi \left( \partial_\nu h^{\mu\nu}-\frac{1}{2} \partial^\mu h^\nu_\nu \right)
\left( \partial^\beta h_{\mu\beta}-\frac{1}{2} \partial_\mu h^\alpha_\alpha \right),
\label{GFT}
\end{equation}
with $\xi$ the gauge parameter, and the Faddeev-Popov determinant is included in the  measure of
integration. The condition of vanishing of the right-hand-side of Eq.~(\ref{VacuumE})
uniquely fixes all coefficients in the  power series expansion of the cosmological constant in
terms of $\hbar$:
\begin{equation}
\Lambda = \sum_{i=0}^\infty \hbar^i \Lambda _i \,.
\label{CCexpanded}
\end{equation}
\noindent
To perform perturbative calculations,  we represent the metric and vielbein fields as sums of the
Minkowskian background and the quantum fields \cite{tHooft:1974toh,Donoghue:2015hwa}
\begin{eqnarray}
g_{\mu\nu} &=& \eta_{\mu\nu}+\kappa h_{\mu\nu} ,\nonumber \\
g^ {\mu\nu} &=& \eta^{\mu\nu}-\kappa h^{\mu\nu}+\kappa^2 h^\mu_\lambda h^{\lambda\nu}-\kappa^3
h^\mu_\lambda h^{\lambda}_{\sigma} h^{\sigma\nu}+\cdots , \nonumber\\
e^a_\mu &=& \delta^a_\mu +\frac{\kappa}{2} \, h^a_\mu -\frac{\kappa^2}{8}\, h_{\mu\rho} h^{a\rho} +\cdots ,
\nonumber \\
e_a^\mu &=& \delta_a^\mu - \frac{\kappa}{2}\, h_{a}^\mu + \frac{3\kappa^2}{8}\, h_{a\rho} h^{\mu\rho}
+ \ldots~ .
\label{tetradexpanded}
\end{eqnarray}
Applying standard quantum field theory  techniques, we obtain the Feynman rules required for the
calculations performed here. These are specified in the appendix when fermion fields are
involved (the other ones are given in the appendix of our earlier paper~\cite{Gegelia:2019fjx}).

\begin{figure}[t]
\begin{center}
\epsfig{file=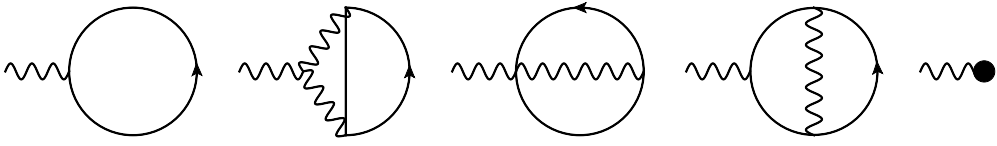,scale=0.8}
\caption{Diagrams contributing to the vacuum expectation value of the graviton field.
The filled circle corresponds to the cosmological constant term. The wiggly and solid lines
represent  gravitons and fermions, respectively.} \label{GR-Tadpole}
\end{center}
\end{figure}

An infinite number of diagrams contribute to the vacuum expectation value of the full
energy-momentum pseudotensor times $(-g)$ at tree order, however, all of them vanish if we
take $\Lambda_0=0$  in Eq.~(\ref{CCexpanded}) \cite{Gegelia:2019fjx}.
Notice that this also removes the mass term from the graviton propagator, corresponding to a ghost degree
of freedom,  at tree order \cite{Veltman:1975vx}.

\begin{figure}[t]
\begin{center}
\epsfig{file=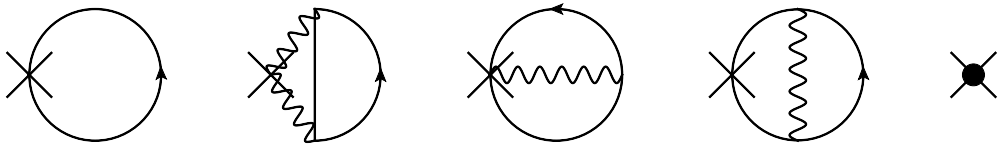,scale=0.8}
\caption{Diagrams contributing to the vacuum expectation value of the energy-momentum  pseudotensor
times $(-g)$. The filled circle corresponds to the cosmological constant term. 
The cross stands for the energy-momentum pseudotensor times $(-g)$, wiggly and solid lines represent
gravitons and  fermions, respectively.}
\label{EMTLoop}
\end{center}
\end{figure}

Next, to obtain the one-loop contributions to the vacuum expectation value of
the full energy-momentum pseudotensor  times $(-g)$, we calculated the corresponding Feynman
diagrams shown in Fig.~\ref{EMTLoop}. By demanding that $\Lambda_1$ cancels this contribution
we obtain (in the calculations of the loop diagrams below we applied dimensional
regularization, with $d$ the dimension of the spacetime, and used the program
FeynCalc \cite{Mertig:1990an,Shtabovenko:2016sxi})
\begin{equation}
\Lambda_1 = \frac{2^{-d} \pi ^{-d/2} \kappa ^2 \mu ^{4-d} m^d \Gamma
   \left(1-\frac{d}{2}\right)}{d}\,,
\end{equation}
with $\mu$ the scale of dimensional regularization and $\Gamma$ is  Euler's $\Gamma$-function.
It is a trivial consequence of the definition of the energy-momentum tensor of the matter fields
that the same value of $\Lambda_1$ cancels the one-loop contribution
to the vacuum expectation value of the graviton field $h_{\mu\nu}$, shown in Fig.~\ref{GR-Tadpole}, and 
consequently, the graviton self-energy at zero momentum, i.e. graviton mass, as a result of
a Ward identity \cite{Burns:2014bva}.
The first non-trivial result is obtained by calculating the two-loop diagrams contributing
to the vacuum expectation value of the full energy-momentum pseudotensor times $(-g)$
shown in Fig.~\ref{EMTLoop} and to the vacuum expectation value of the gravitational field shown
in Fig.~\ref{GR-Tadpole}. The same value
\begin{equation}
\Lambda_2 = -\frac{2^{-2 d-7} d^3 \pi ^{1-d} \kappa ^4 \mu ^{8-2 d} m^{2 d-2}
   \csc \left(\frac{\pi  d}{2}\right) \Gamma
   \left(-\frac{d}{2}\right)}{(d-2) \Gamma
   \left(\frac{d}{2}\right)}
\label{Lambda2}
\end{equation}
cancels both quantities. Here, $\csc$ is the cosecans. To check the  reliability of the
obtained results  we also calculated the two-loop contributions to the graviton self-energy 
and checked that the same value of $\Lambda_2$ ensures that the graviton remains 
massless in agreement with the Ward identity \cite{Burns:2014bva} (we do not give the expressions
of the  Feynman rules needed for the calculation of the graviton self-energy due to their huge size).

\section{QCD contribution to the vacuum energy}
\label{QCDC}

In the framework of general relativity coupled to the Standard Model, the contribution to the
cosmological constant originating from the shift in the vacuum energy due to explicit breaking
of chiral symmetry of QCD can be calculated with great accuracy~\cite{Donoghue:2016tjk}. 
Consider the two-flavour QCD Lagrangian of massless up and down quarks with external scalar
and pseudoscalar currents $s(x)$ and $p(x)$, respectively, 
\begin{equation}
{\cal  L} = 
-\frac{1}{4} F^a_{\mu\nu} F^{a\mu\nu} +i\,\bar \psi \gamma_\alpha D^\alpha \psi 
- \bar \psi (s-i\,\gamma_5 p)\psi\,,
\label{QCDLagr}
\end{equation}
where $\psi=(\psi_u,\psi_d)^T$ is a doublet comprising  the up and down quark fields.
For simplicity, we only consider the two-flavor case here, the extension to three
flavors (adding the strange quark)  is straightforward.
The Lagrangian of Eq.~(\ref{QCDLagr}) is invariant under $SU(2)_L\times SU(2)_R$ chiral
symmetry transformations 
\begin{eqnarray}
\frac{1}{2} (1-\gamma_5) \psi \to L \, \frac{1}{2} (1-\gamma_5) \psi\,,\ \ \
\frac{1}{2} (1+\gamma_5) \psi \to R \, \frac{1}{2} (1+\gamma_5) \psi\,,\ \ \
 (s+i\,p) \to L \, (s+i \,p) R^\dagger\,,
\label{transs}
\end{eqnarray}
with $L$ and $R$  elements of $SU(2)_L$ and $SU(2)_R$, respectively. Massless QCD undergoes
spontaneous symmetry breaking with pions appearing as Goldstone bosons.  
The corresponding low-energy effective Lagrangian is given as an expansion in chiral orders,
also taking into account the anomaly of the singlet axial current  \cite{Gasser:1984yg}. 
The lowest order effective Lagrangian has the form  
\begin{equation}
  {\cal L}_2 = \frac{F_\pi^2}{4}\,{\rm Tr}(\partial_\mu U\partial^\mu U^\dagger) +
  \frac{F_\pi^2}{4}\,{\rm Tr}(\chi U^\dagger+U\chi^\dagger)~,
\label{LagrL2}
\end{equation}
where ${\rm Tr}$ denotes the trace in flavor (isospin) space and the matrix-valued field $U$ is given
in terms of the pion fields $\pi^a$ ($a=1,2,3$) as  
\begin{equation}
U =\exp \left(\frac{i\, \tau^a \pi^a}{F_\pi}\right)\,,
\end{equation}
with $\tau^a$ the Pauli matrices and $\chi = 2 B_0 (s+i p)$. Here $B_0$ is a constant of dimension [mass] 
related to the vacuum expectation value of the scalar quark condensate and $F_\pi$ is the pion decay
constant (in the chiral limit).
The Lagrangian of Eq.~(\ref{LagrL2}) is invariant under chiral transformations
\begin{equation}
U\to L U R^\dagger\,, \ \  (s+i\,p)\to L (s+i\,p) R^\dagger .
\label{chtransL2}
\end{equation}
The effective field theory corresponding to  QCD is obtained by substituting the external sources as follows
\begin{equation}
s = \left[
\begin{array}{cc}
m_u            &  0 \\
0  &  m_d           
\end{array}
\right] , \ \ p=0\,.
\label{exts}
\end{equation}
As the quark masses explicitly break the chiral symmetry, the pions obtain a small mass 
to leading order in the chiral expansion (much smaller than any other hadron mass) 
\begin{equation}
M_\pi^2 = B_0 (m_u+m_d) +{\cal O}(m_q^2) ~,
\label{pionmass}
\end{equation}
where $m_q$ denotes any of the light quark masses.
Further, the effective Lagrangian generates a 
tree-order contribution to the vacuum energy  \cite{Donoghue:2016tjk} 
\begin{equation}
\Lambda_m = -\langle 0| {\cal L}_2|0\rangle= -F_\pi^2 B_0 (m_u+m_d) =-F_\pi^2 M_\pi^2~.
\label{lambdaeft}
\end{equation}
There is no other term linear in the quark
masses in the chiral effective Lagrangian which could compensate the contribution of Eq.~(\ref{lambdaeft}),
e.g. a term like ${\rm Tr}(\chi +\chi^\dagger)$ would contribute to
the vacuum energy but it violates the chiral symmetry.

It is argued in Ref.~\cite{Donoghue:2016tjk} that to cancel the contribution to the vacuum energy
given in Eq.~(\ref{lambdaeft}) one needs to adjust numerically the cosmological constant term 
in the EFT of pions interacting with gravitation where it is one of the parameters of the
effective Lagrangian. Evaluating Eq.~(\ref{lambdaeft}),  one obtains that the chiral symmetry breaking term of
 QCD gives a large contribution to  the vacuum energy
\begin{equation}
\Lambda_m =1.5\times 10^8 \, {\rm MeV}^4 = 0.63\times10^ {43}\Lambda_{\rm exp},
\end{equation}
where $\Lambda_{\rm exp}=2.4\times10^{-47}$ ${\rm GeV}^4$ is the observed value of the cosmological constant  \cite{Tanabashi:2018oca}. 
It is stated in Ref.~\cite{Donoghue:2016tjk} that: ``Because of the large multiplier, if one holds all the other parameters of the Standard Model fixed, a change of the up quark mass in its forty-first 
digit would produce a change in $\Lambda$ outside the anthropically allowed range. ... because the calculation is so well controlled, it illustrates the degree of fine-tuning required as well as the 
futility of thinking that some feature of the Standard Model could lead to a vanishing contribution to $\Lambda$.'' Below we critically examine  this statement.

\begin{figure}[t]
\begin{center}
\epsfig{file=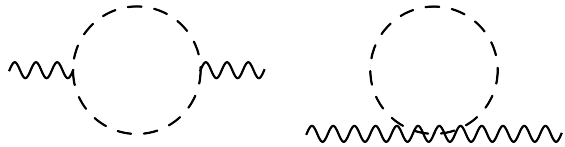,scale=0.5}
\caption{One-loop diagrams with pions contributing to the graviton self-energy. Wiggly and dashed lines represent gravitons
 and  pions, respectively. }
\label{fig2}
\end{center}
\end{figure}

It is straightforward to construct a low-energy EFT of pions including the interaction with  the gravitational field. The corresponding action has the form  
\begin{equation}
S = S_{\rm gr}(g) + 
\int d^4x \sqrt{-g}\, \left[ - F_\pi^2 M_\pi^2 + \frac{1}{2} \, g^{\mu\nu} \partial_\mu \pi^a \, \partial_\nu \pi^a  -  \frac{1}{2} \, M_\pi^2 \pi^a \pi^a + {\cal O}(\pi^4) \right] ~.  
\label{ChPT_action}
\end{equation}
Following the logic of the previous section to cancel the tree order contribution of the pions to the vacuum energy we need to take 
$\Lambda_0 =  F_\pi^2 M_\pi^2\,$
in the expansion of Eq.~(\ref{CCexpanded}). This value of $\Lambda_0$ exactly cancels also the graviton mass generated by Eq.~(\ref{ChPT_action}) at tree order.
At one-loop order there are two diagrams, shown in Fig.~\ref{fig2}, contributing to the graviton self-energy generated by the effective Lagrangian of Eq.~(\ref{ChPT_action}). 
By demanding that the order $\hbar$ term in Eq.~(\ref{CCexpanded}) exactly cancels the contribution of these two one-loop diagrams  for $p^2=0$ ($p$ is the four-momentum of the graviton) we obtain
\begin{equation}
\Lambda_1 = \frac{3 \kappa ^2
   \left(2 M_{\pi }^2 \,A_0 \left(M_{\pi
   }^2\right)+M_{\pi }^4\right)}{512 \pi
   ^2}\,,
\label{Lambda1}
\end{equation}
where using  dimensional regularization for the loop integral we have
\begin{eqnarray}
A_0(M^2) & = &\frac{(2 \pi)^{4-d}\mu^{4-d}}{i\,\pi^2}\,\int
\frac{d^dk}{k^2-M^2+i 0^+}\,.
\label{Bointegral}
\end{eqnarray}
Thus, in full agreement with the results of Ref.~\cite{Burns:2014bva} to have a self-consistent EFT the cosmological constant as a parameter of this 
theory has to be a fixed function of the light quark masses (or equivalently, of the pion mass).  
This condition imposed on the cosmological constant requires that the cosmological constant term exactly cancels contributions of matter fields to vacuum energy analytically, for any 
values of the masses and couplings. Such a condition invalidates the considerations of Ref.~\cite{Donoghue:2016tjk} about the 
numerical fine tuning briefly recapitulated above.  
However,   chiral invariance of the low-energy effective Lagrangian of pions interacting with gravitation does not allow a quark mass dependent cosmological constant term, 
see the discussion after Eq.~(\ref{lambdaeft}).

Thus, on the one hand the consistency condition of the EFT of general relativity requires that the cosmological constant term is a given fixed function of the light quark masses 
and on the other hand the chiral symmetry of QCD does not allow such a term.
The solution to this apparent problem is that the chiral symmetry of the QCD Lagrangian with external sources is not an exact symmetry of the full theory including the gravity.

Let us have a closer look at the action of general relativity given by Eq.~(\ref{action}).  According to Ref.~\cite{Burns:2014bva} the cosmological constant term has to be a fixed 
function of other parameters of the theory, i.e. $\Lambda\equiv \Lambda(m_u,m_d,g,e,\ldots)$, where $e$ is the electromagnetic coupling and the ellipsis  stands for other parameters of the effective theory.
The Lagrangian ${\cal L}_{\rm m}(g,\psi)$ at leading order coincides with the Lagrangian of the Standard Model, i.e. QCD plus the electroweak theory, taken in a non-flat metric field.
To obtain the leading order Lagrangian of the strong interaction we ``switch off'' all other interactions and drop interaction terms with negative mass dimensions (i.e. higher-order ``non-renormalizable" interactions). 
To  ``switch off" gravity we approximate the metric field $g^{\mu\nu}$ by the constant Minkowski metric and for the electroweak interaction we put the corresponding couplings equal to zero. 
This leaves us with the following Lagrangian (for two flavours of quarks) 
\begin{equation}
{\cal  L} = -\frac{1}{4} F^a_{\mu\nu} F^{a\mu\nu} + \bar \psi \left( i\, \gamma_\alpha D^\alpha -{\cal M}\right) \psi + L_0(m_u,m_d,g)\,,
\label{QCDLagr2}
\end{equation}
where we have denoted $-4 \Lambda(m_u,m_d,g,0,0,\cdots )/\kappa^2$  by $L_0(m_u,m_d, g)$. 
This term does not contradict to any {\it  physical symmetries}, however, it is not usually included 
in the QCD Lagrangian because it does not contribute in  physical quantities when  gravity is not taken into account. 

The Lagrangian of Eq.~(\ref{QCDLagr2})  leads to the following contribution to the vacuum energy \cite{Donoghue:2016tjk1}
\begin{equation}
\Lambda_m= \langle 0| m_u \bar\psi_u \psi_u+m_d \bar\psi_d\psi_d|0\rangle - \langle 0| L_0(m_u,m_d,g) |0\rangle =-F_\pi^2 M_\pi^2 - L_0(m_u,m_d,g) .
\label{lambda1}
\end{equation}
By taking the yet unspecified constant term of the QCD Lagrangian as $L_0(m_u,m_d,g)=-F_\pi^2 M_\pi^2+{\cal O}(M_\pi^4)=-F_\pi^2 B_0 (m_u+m_d)+{\cal O}(m_q^2)$ and substituting in 
Eq.~(\ref{lambda1})
we obtain  for the contribution to the vacuum energy $\Lambda_m= 0 +{\cal O}(m_q^2) $. 
By adjusting the terms of higher orders in light quark masses $m_q$ in $L_0(m_u,m_d,g)$ we can achieve that the QCD contribution to the vacuum energy $\Lambda_m$ exactly 
vanishes for any values of the quark masses.

While cancelling the standard QCD contribution to the vacuum energy, the addition of the constant $L_0$ term to the 
Lagrangian does not affect the construction 
of the low-energy effective field theory which proceeds in exact analogy to  Ref.~\cite{Gasser:1984yg} by starting with
the following Lagrangian with external sources 
\begin{equation}
{\cal L}_{\rm ext} = -\frac{1}{4} F^a_{\mu\nu} F^{a\mu\nu} +i\,\bar \psi \gamma_\alpha D^\alpha \psi + \bar \psi\gamma^\mu(v_\mu + a_\mu\gamma_5) \psi
- \bar \psi (s-i\,\gamma_5 p)\psi + L_0(m_u,m_d,g),
\label{QCDLagrE}
\end{equation}
where the external sources  $v_\mu(x)$, $a_\mu(x)$,  $s(x)$ and $p(x)$ are Hermitean, color neutral matrices in flavour space and $s(x)={\cal M}+\cdots $ incorporates the 
quark mass term. 
Greens functions of scalar, pseudo-scalar, vector and axial vector currents  are generated by the vacuum-to-vacuum transition amplitude 
\begin{equation}
\langle 0_{\rm out} |0_{\rm in} \rangle_{v,a,s,p} = e^{i Z[v,a,s,p]}=\frac{\int {\cal D} A {\cal D} q \,e^{i\int d^4 x {\cal L}_{\rm ext}(x)}}{\int {\cal D} A {\cal D} q \,e^{i\int d^4 x {\cal L}(x)}}.
\label{genfunct}
\end{equation}
The generating functional $Z[v,a,s,p]$ clearly does not depend on $L_0$ and therefore the construction of the low-energy EFT, namely chiral perturbation theory, is 
exactly the same as in Ref.~\cite{Gasser:1984yg} exploiting the fact that 
the Lagrangian ${\cal L}_{\rm ext}$ { without the $L_0$ term} is invariant under local transformations 
\begin{equation}
\psi(x)\to \left[ \frac{1}{2} (1+\gamma_5) R(x) +\frac{1}{2} (1-\gamma_5) L(x) \right] \psi(x)
\label{qtranss}
\end{equation}
provided that the external sources transform as follows,
\begin{eqnarray}
v_\mu' + a_\mu' &=& R(v_\mu+a_\mu )R^\dagger +i R\partial_\mu R^\dagger ,\nonumber\\
v_\mu' - a_\mu' &=& L(v_\mu-a_\mu )L^\dagger +i L\partial_\mu L^\dagger,\nonumber\\
s' +i  p' &=& R(s+i p) L.
\label{extranss}
\end{eqnarray}
We conclude that the presence in the QCD Lagrangian of the $L_0(m_u,m_d,g)$ term, which is nothing else then a cosmological constant, does not contradict to any physically relevant symmetries of QCD.

\section{Summary}
\label{summary}

By demanding  the presence of a massless graviton (instead of a massive spin-two ghost)
in the spectrum of the perturbative EFT of general relativity in flat Minkowski background the cosmological constant term is uniquely fixed
 as a function of all other parameters of the theory \cite{Burns:2014bva}. 
We argue that if there is any physical reason for choosing a fixed value of the cosmological constant then it
must be the condition of vanishing of the vacuum energy. 
In our recent paper \cite{Gegelia:2019fjx} we calculated the vacuum expectation value of the full four-momentum of the matter and gravitational
fields at two-loop order in a simplified version of the Abelian model with spontaneous symmetry breaking
considered also in Ref.~\cite{Burns:2014bva}. 
We obtained that as a result of a non-trivial cancellation between different diagrams the requirement of the vanishing vacuum energy leads to the conditions of
Ref.~\cite{Burns:2014bva}. While in Ref.~\cite{Gegelia:2019fjx} we included only a scalar and vectors as the matter fields, in the current work we considered the contributions of fermions and obtained 
similar results. In particular, the value of the cosmological constant which cancels the two-loop fermion contribution to the vacuum energy also eliminates the vacuum expectation value of the graviton field and the 
massive ghost, thus leading to a self-consistent EFT of general relativity in Minkowski background. 
Being aware of  the non-existence of a commonly accepted expression of the energy-momentum tensor for the
gravitational field (see, e.g., Refs.~\cite{Babak:1999dc,Butcher:2008rf,Szabados:2009eka,Butcher:2010ja,Butcher:2012th}), 
we used the definition of the energy-momentum pseudotensor and the full four-momentum given in
the classic textbook by Landau and Lifshitz~\cite{Landau:1982dva}. 

While we are still unable to give a general argument,
based on our two-loop order results in an EFT  of matter and gravitational fields on flat Minkowski background 
we expect that by demanding that the vacuum energy should be vanishing to all orders 
we obtain a self-consistent perturbative EFT of gravitation coupled to the fields of the Standard Model. 

The results of Refs.~\cite{Burns:2014bva,Gegelia:2019fjx}  and of the current work resolve the issue of the 
fine-tuning of the QCD contribution in the vacuum energy addressed in Ref.~\cite{Donoghue:2016tjk}. In particular, there is no numerical fine-tuning but rather 
the cosmological constant, as a function of the parameters of QCD, exactly cancels the QCD contribution to the vacuum energy. 
However, this solution of the problem seems to be incompatible with the chiral symmetry of QCD.
A closer examination, however, reveals that there is no contradiction with any symmetries of QCD with observable physical consequences.

\medskip

\acknowledgments

The work of JG was supported in part by BMBF (Grant No. 05P18PCFP1), and by the
Georgian Shota Rustaveli National Science Foundation (Grant No. FR17-354).
The work of UGM was supported in part by provided by Deutsche
Forschungsgemeinschaft (DFG)  through funds provided to the Sino-German CRC~110
``Symmetries and the Emergence of Structure in QCD" (Grant No. TRR110),  by
the Chinese Academy of Sciences (CAS) through a President's International Fellowship
Initiative (PIFI) (Grant No. 2018DM0034) and by the VolkswagenStiftung (Grant No. 93562).

\appendix 

\section{Feynman rules}


Below we give Feynman rules involving fermions used in the calculation of the vacuum
expectation values of the graviton field and the energy-momentum tensor multiplied with $(-g)$.
The other Feynman rules are given in the appendix of Ref.~\cite{Gegelia:2019fjx}.

\medskip

\noindent
Propagators:

\begin{itemize}

\item[a)]
Fermion propagator with  momentum $p$:
\begin{equation}
 \frac{ i }{ \slashed p-m+i
   \epsilon }\,.
\label{sPr}
\end{equation}

\end{itemize}

\medskip

\noindent
Vertices (all momenta in all vertices are incoming): 
\begin{itemize}
\item[a)]
  Graviton with indices $(\mu,\nu)$ - incoming fermion with 
incoming momentum $p_1$  and outgoing fermion with incoming momentum $p_2$:
\begin{eqnarray}
-\frac{1}{8} i \kappa  \left[2 g^{\mu \nu }
   \left(2 m-\gamma \cdot p_1+\gamma \cdot
   p_2\right)+\gamma ^{\mu } \left(p_1{}^{\nu
   }-p_2{}^{\nu }\right)+\gamma ^{\nu }
   \left(p_1{}^{\mu }-p_2{}^{\mu }\right)\right];
\label{hFF}
\end{eqnarray}

\item[b)]
Gravitons with (Lorentz indices, momentum)
  combinations $(\mu,\nu, k_1)$  and $(\alpha,\beta, k_2)$ - incoming fermion with 
incoming momentum $p_1$  and outgoing fermion with incoming momentum $p_2$:
\begin{eqnarray}
&& \frac{i \kappa ^2}{256}  \left\{ 2 \gamma ^{\beta
   }.\gamma ^{\mu }.\gamma ^{\nu } k_1{}^{\alpha
   }+2 \gamma ^{\beta }.\gamma ^{\nu }.\gamma
   ^{\mu } k_1{}^{\alpha }  -2 \gamma ^{\mu
   }.\gamma ^{\nu }.\gamma ^{\beta }
   k_1{}^{\alpha }-2 \gamma ^{\nu }.\gamma ^{\mu
   }.\gamma ^{\beta } k_1{}^{\alpha }+2 \gamma
   ^{\alpha }.\gamma ^{\mu }.\gamma ^{\nu }
   k_1{}^{\beta }+2 \gamma ^{\alpha }.\gamma
   ^{\nu }.\gamma ^{\mu } k_1{}^{\beta }  
   \right.  \nonumber\\
    && \left. 
    -2 \gamma
   ^{\mu }.\gamma ^{\nu }.\gamma ^{\alpha }
   k_1{}^{\beta }-2 \gamma ^{\nu }.\gamma ^{\mu
   }.\gamma ^{\alpha } k_1{}^{\beta }-2 \gamma
   ^{\alpha }.\gamma ^{\beta }.\gamma ^{\nu }
   k_2{}^{\mu }-2 \gamma ^{\beta }.\gamma
   ^{\alpha }.\gamma ^{\nu } k_2{}^{\mu }+2
   \gamma ^{\nu }.\gamma ^{\alpha }.\gamma
   ^{\beta } k_2{}^{\mu }+2 \gamma ^{\nu }.\gamma
   ^{\beta }.\gamma ^{\alpha } k_2{}^{\mu }
   \right.  \nonumber\\
    && \left.
   -2
   \gamma ^{\alpha }.\gamma ^{\beta }.\gamma
   ^{\mu } k_2{}^{\nu }-2 \gamma ^{\beta }.\gamma
   ^{\alpha }.\gamma ^{\mu } k_2{}^{\nu }+2
   \gamma ^{\mu }.\gamma ^{\alpha }.\gamma
   ^{\beta } k_2{}^{\nu }+2 \gamma ^{\mu }.\gamma
   ^{\beta }.\gamma ^{\alpha } k_2{}^{\nu }+4
   \gamma ^{\mu }.\gamma ^{\nu }.\left(\gamma
   \cdot k_1\right) g^{\alpha \beta }+4 \gamma
   ^{\nu }.\gamma ^{\mu }.\left(\gamma \cdot
   k_1\right) g^{\alpha \beta }
   \right.  \nonumber\\
    && \left.
   -4 \left(\gamma
   \cdot k_1\right).\gamma ^{\mu }.\gamma ^{\nu }
   g^{\alpha \beta }
   -4 \left(\gamma \cdot
   k_1\right).\gamma ^{\nu }.\gamma ^{\mu }
   g^{\alpha \beta }-3 \gamma ^{\beta }.\gamma
   ^{\nu }.\left(\gamma \cdot k_1\right)
   g^{\alpha \mu }-\gamma ^{\beta }.\gamma ^{\nu
   }.\left(\gamma \cdot k_2\right) g^{\alpha \mu
   }-\gamma ^{\nu }.\gamma ^{\beta }.\left(\gamma
   \cdot k_1\right) g^{\alpha \mu }
   \right.  \nonumber\\
    && \left.
      -3 \gamma
   ^{\nu }.\gamma ^{\beta }.\left(\gamma \cdot
   k_2\right) g^{\alpha \mu }+\left(\gamma \cdot
   k_1\right).\gamma ^{\beta }.\gamma ^{\nu }
   g^{\alpha \mu }+3 \left(\gamma \cdot
   k_1\right).\gamma ^{\nu }.\gamma ^{\beta }
   g^{\alpha \mu }+3 \left(\gamma \cdot
   k_2\right).\gamma ^{\beta }.\gamma ^{\nu }
   g^{\alpha \mu }+\left(\gamma \cdot
   k_2\right).\gamma ^{\nu }.\gamma ^{\beta }
   g^{\alpha \mu }
   \right.  \nonumber\\
    && \left.
   -3 \gamma ^{\beta }.\gamma
   ^{\mu }.\left(\gamma \cdot k_1\right)
   g^{\alpha \nu }-\gamma ^{\beta }.\gamma ^{\mu
   }.\left(\gamma \cdot k_2\right) g^{\alpha \nu
   }-\gamma ^{\mu }.\gamma ^{\beta }.\left(\gamma
   \cdot k_1\right) g^{\alpha \nu }-3 \gamma
   ^{\mu }.\gamma ^{\beta }.\left(\gamma \cdot
   k_2\right) g^{\alpha \nu }+\left(\gamma \cdot
   k_1\right).\gamma ^{\beta }.\gamma ^{\mu }
   g^{\alpha \nu }
   \right.  \nonumber\\
    && \left.   
   +3 \left(\gamma \cdot
   k_1\right).\gamma ^{\mu }.\gamma ^{\beta }
   g^{\alpha \nu }+3 \left(\gamma \cdot
   k_2\right).\gamma ^{\beta }.\gamma ^{\mu }
   g^{\alpha \nu }+\left(\gamma \cdot
   k_2\right).\gamma ^{\mu }.\gamma ^{\beta }
   g^{\alpha \nu }-3 \gamma ^{\alpha }.\gamma
   ^{\nu }.\left(\gamma \cdot k_1\right) g^{\beta
   \mu }-\gamma ^{\alpha }.\gamma ^{\nu
   }.\left(\gamma \cdot k_2\right) g^{\beta \mu
   }
   \right.  \nonumber\\
    && \left.
   -\gamma ^{\nu }.\gamma ^{\alpha
   }.\left(\gamma \cdot k_1\right) g^{\beta \mu
   }-3 \gamma ^{\nu }.\gamma ^{\alpha
   }.\left(\gamma \cdot k_2\right) g^{\beta \mu
   }+\left(\gamma \cdot k_1\right).\gamma
   ^{\alpha }.\gamma ^{\nu } g^{\beta \mu }+3
   \left(\gamma \cdot k_1\right).\gamma ^{\nu
   }.\gamma ^{\alpha } g^{\beta \mu }+3
   \left(\gamma \cdot k_2\right).\gamma ^{\alpha
   }.\gamma ^{\nu } g^{\beta \mu }
   \right.  \nonumber\\
    && \left.
   +\left(\gamma
   \cdot k_2\right).\gamma ^{\nu }.\gamma
   ^{\alpha } g^{\beta \mu }-3 \gamma ^{\alpha
   }.\gamma ^{\mu }.\left(\gamma \cdot k_1\right)
   g^{\beta \nu }-\gamma ^{\alpha }.\gamma ^{\mu
   }.\left(\gamma \cdot k_2\right) g^{\beta \nu
   }-\gamma ^{\mu }.\gamma ^{\alpha
   }.\left(\gamma \cdot k_1\right) g^{\beta \nu
   }-3 \gamma ^{\mu }.\gamma ^{\alpha
   }.\left(\gamma \cdot k_2\right) g^{\beta \nu
   }
   \right.  \nonumber\\
    && \left.
   +\left(\gamma \cdot k_1\right).\gamma
   ^{\alpha }.\gamma ^{\mu } g^{\beta \nu }+3
   \left(\gamma \cdot k_1\right).\gamma ^{\mu
   }.\gamma ^{\alpha } g^{\beta \nu }+3
   \left(\gamma \cdot k_2\right).\gamma ^{\alpha
   }.\gamma ^{\mu } g^{\beta \nu }+\left(\gamma
   \cdot k_2\right).\gamma ^{\mu }.\gamma
   ^{\alpha } g^{\beta \nu }+4 \gamma ^{\alpha
   }.\gamma ^{\beta }.\left(\gamma \cdot
   k_2\right) g^{\mu \nu }
   \right.  \nonumber\\
    && \left.
   +4 \gamma ^{\beta
   }.\gamma ^{\alpha }.\left(\gamma \cdot
   k_2\right) g^{\mu \nu }-4 \left(\gamma \cdot
   k_2\right).\gamma ^{\alpha }.\gamma ^{\beta }
   g^{\mu \nu }-4 \left(\gamma \cdot
   k_2\right).\gamma ^{\beta }.\gamma ^{\alpha }
   g^{\mu \nu }+4 \left(-4 p_1{}^{\nu } \gamma
   ^{\mu } g^{\alpha \beta }+4 p_2{}^{\nu }
   \gamma ^{\mu } g^{\alpha \beta }
   \right.
    \right.  \nonumber\\
    && \left.\left.
   -4 p_1{}^{\mu
   } \gamma ^{\nu } g^{\alpha \beta }+4
   p_2{}^{\mu } \gamma ^{\nu } g^{\alpha \beta
   }-16 m g^{\mu \nu } g^{\alpha \beta }+8 \gamma
   \cdot p_1 g^{\mu \nu } g^{\alpha \beta }-8
   \gamma \cdot p_2 g^{\mu \nu } g^{\alpha \beta
   }
   +3 p_1{}^{\beta } \gamma ^{\nu } g^{\alpha
   \mu }-3 p_2{}^{\beta } \gamma ^{\nu }
   g^{\alpha \mu }
   \right.\right.  \nonumber\\
    && \left.\left.
   +3 p_1{}^{\beta } \gamma ^{\mu
   } g^{\alpha \nu }-3 p_2{}^{\beta } \gamma
   ^{\mu } g^{\alpha \nu }+\left(3
   \left(p_1{}^{\nu }-p_2{}^{\nu }\right) \gamma
   ^{\alpha }+3 \left(p_1{}^{\alpha
   }-p_2{}^{\alpha }\right) \gamma ^{\nu }+8
   \left(2 m-\gamma \cdot p_1+\gamma \cdot
   p_2\right) g^{\alpha \nu }\right) g^{\beta \mu
   }
   \right.\right.  \nonumber\\
    && \left.\left.
   +3 p_1{}^{\mu } \gamma ^{\alpha } g^{\beta
   \nu }-3 p_2{}^{\mu } \gamma ^{\alpha }
   g^{\beta \nu }+3 p_1{}^{\alpha } \gamma ^{\mu
   } g^{\beta \nu }-3 p_2{}^{\alpha } \gamma
   ^{\mu } g^{\beta \nu }+16 m g^{\alpha \mu }
   g^{\beta \nu }-8 \gamma \cdot p_1 g^{\alpha
   \mu } g^{\beta \nu }+8 \gamma \cdot p_2
   g^{\alpha \mu } g^{\beta \nu }
   \right.\right.  \nonumber\\
    && \left.\left.
   -4 p_1{}^{\beta
   } \gamma ^{\alpha } g^{\mu \nu }+4
   p_2{}^{\beta } \gamma ^{\alpha } g^{\mu \nu
   }+\gamma ^{\beta } \left(3 p_1{}^{\nu }
   g^{\alpha \mu }-3 p_2{}^{\nu } g^{\alpha \mu
   }+3 p_1{}^{\mu } g^{\alpha \nu }-3 p_2{}^{\mu
   } g^{\alpha \nu }-4 p_1{}^{\alpha } g^{\mu \nu
   }+4 p_2{}^{\alpha } g^{\mu \nu
   }\right)\right)\right\};
\label{hhFF}
\end{eqnarray}

\item[c)]
Energy-momentum tensor with indices $(\mu,\nu)$ - 
incoming fermion with 
incoming momentum $p_1$  and outgoing fermion with incoming momentum $p_2$:
\begin{equation}
\frac{1}{2} \left(\gamma ^{\mu } p_1{}^{\nu
   }+\gamma ^{\nu } p_1{}^{\mu }\right) ;
\label{TFF}
\end{equation}

\item[d)]
  Energy-momentum tensor with indices $(\mu,\nu)$ - graviton with (Lorentz indices, momentum)
  combination $(\alpha,\beta,k_1)$ - incoming fermion with 
incoming momentum $p_1$  and outgoing fermion with incoming momentum $p_2$:
\begin{eqnarray}
&& \frac{1}{32} \kappa  \left\{ -\gamma ^{\beta
   }.\left(\gamma \cdot k_1\right).\gamma ^{\nu }
   g^{\alpha \mu }-\gamma ^{\nu }.\gamma ^{\beta
   }.\left(\gamma \cdot k_1\right) g^{\alpha \mu
   }+\gamma ^{\nu }.\left(\gamma \cdot
   k_1\right).\gamma ^{\beta } g^{\alpha \mu
   }+\left(\gamma \cdot k_1\right).\gamma ^{\beta
   }.\gamma ^{\nu } g^{\alpha \mu }-\gamma
   ^{\beta }.\left(\gamma \cdot k_1\right).\gamma
   ^{\mu } g^{\alpha \nu }  
   \right.  \nonumber\\
&&   \left.   -\gamma ^{\mu }.\gamma
   ^{\beta }.\left(\gamma \cdot k_1\right)
   g^{\alpha \nu }+\gamma ^{\mu }.\left(\gamma
   \cdot k_1\right).\gamma ^{\beta } g^{\alpha
   \nu }+\left(\gamma \cdot k_1\right).\gamma
   ^{\beta }.\gamma ^{\mu } g^{\alpha \nu
   }-\gamma ^{\alpha }.\left(\gamma \cdot
   k_1\right).\gamma ^{\nu } g^{\beta \mu
   }-\gamma ^{\nu }.\gamma ^{\alpha
   }.\left(\gamma \cdot k_1\right) g^{\beta \mu
   }
   \right.  \nonumber\\
&&   \left. 
   +\gamma ^{\nu }.\left(\gamma \cdot
   k_1\right).\gamma ^{\alpha } g^{\beta \mu
   }+\left(\gamma \cdot k_1\right).\gamma
   ^{\alpha }.\gamma ^{\nu } g^{\beta \mu
   }-\gamma ^{\alpha }.\left(\gamma \cdot
   k_1\right).\gamma ^{\mu } g^{\beta \nu
   }-\gamma ^{\mu }.\gamma ^{\alpha
   }.\left(\gamma \cdot k_1\right) g^{\beta \nu
   }+\gamma ^{\mu }.\left(\gamma \cdot
   k_1\right).\gamma ^{\alpha } g^{\beta \nu
   }
   \right.  \nonumber\\
&&   \left. 
   +\left(\gamma \cdot k_1\right).\gamma
   ^{\alpha }.\gamma ^{\mu } g^{\beta \nu }-2
   \left(\gamma ^{\beta } g^{\alpha \nu }
   p_1{}^{\mu }+\gamma ^{\alpha } g^{\beta \nu }
   p_1{}^{\mu }-\gamma ^{\beta } g^{\alpha \nu }
   p_2{}^{\mu }-\gamma ^{\alpha } g^{\beta \nu }
   p_2{}^{\mu }+2 \gamma ^{\nu }
   \left(p_1{}^{\alpha } g^{\beta \mu
   }-p_2{}^{\alpha } g^{\beta \mu }+g^{\alpha \mu
   } p_1{}^{\beta }
   \right. \right. \right. \nonumber\\
&&   \left. \left. \left.
   -g^{\alpha \mu } p_2{}^{\beta
   }-2 g^{\alpha \beta } p_1{}^{\mu }+2 g^{\alpha
   \beta } p_2{}^{\mu }\right)+\gamma ^{\beta }
   g^{\alpha \mu } p_1{}^{\nu }+\gamma ^{\alpha }
   g^{\beta \mu } p_1{}^{\nu }-\gamma ^{\beta }
   g^{\alpha \mu } p_2{}^{\nu }-\gamma ^{\alpha }
   g^{\beta \mu } p_2{}^{\nu } 
   \right. \right.  \nonumber\\
&&   \left. \left.
   +2 \gamma ^{\mu }
   \left(p_1{}^{\alpha } g^{\beta \nu
   }-p_2{}^{\alpha } g^{\beta \nu }+g^{\alpha \nu
   } p_1{}^{\beta }-g^{\alpha \nu } p_2{}^{\beta
   }-2 g^{\alpha \beta } p_1{}^{\nu }+2 g^{\alpha
   \beta } p_2{}^{\nu }\right)\right)\right\};
\label{ThFF}
\end{eqnarray}

\end{itemize}

\medskip

\noindent 
The two-loop master integral appearing in the results of various two-loop calculations is :
\begin{eqnarray}
  && \int\frac{d^d k_1 d^d k_2}{(2 \pi)^{2d}} \frac{1}{(k_1^2-M^2+i \epsilon)^\alpha
    (k_2^2-M^2+i \epsilon)^\beta ((k_1-k_2)^2+i \epsilon)^\gamma} = \nonumber \\
&& \qquad \qquad \ \ \ \frac{ i^{2-2 \alpha -2 \beta -2
   \gamma} M^{2 (d -\alpha -\beta
   -\gamma)} \Gamma \left(\frac{d}{2}-\gamma \right) \Gamma
   \left(\alpha +\gamma -\frac{d}{2} \right) \Gamma
   \left(\beta +\gamma -\frac{d}{2} \right)  \Gamma (\alpha +\beta +\gamma -d )}{(4\pi)^d\Gamma (\alpha
   ) \Gamma (\beta ) \Gamma \left(\frac{d}{2}\right) \Gamma
   (\alpha +\beta +2 \gamma -d)} \,.
\label{int2loop}
\end{eqnarray}

\end{document}